%% file: main.tex
\documentclass[conference]{IEEEtran}
\IEEEoverridecommandlockouts
\usepackage{cite}
\usepackage{amsmath,amssymb,amsfonts}
\usepackage[utf8]{inputenc}

\usepackage{algorithmic}
\usepackage{graphicx}
\usepackage{textcomp}
\usepackage{xcolor}
\usepackage{url}
\usepackage{tikz,pgfplots}
\def\BibTeX{{\rm B\kern-.05em{\sc i\kern-.025em b}\kern-.08em
    T\kern-.1667em\lower.7ex\hbox{E}\kern-.125emX}}
    
\makeatletter
\pgfdeclareshape{document}{
\inheritsavedanchors[from=rectangle] 
\inheritanchorborder[from=rectangle]
\inheritanchor[from=rectangle]{center}
\inheritanchor[from=rectangle]{north}
\inheritanchor[from=rectangle]{south}
\inheritanchor[from=rectangle]{west}
\inheritanchor[from=rectangle]{east}
\backgroundpath{
\southwest \pgf@xa=\pgf@x \pgf@ya=\pgf@y
\northeast \pgf@xb=\pgf@x \pgf@yb=\pgf@y
\pgf@xc=\pgf@xb \advance\pgf@xc by-7.5pt 
\pgf@yc=\pgf@yb \advance\pgf@yc by-7.5pt
\pgfpathmoveto{\pgfpoint{\pgf@xa}{\pgf@ya}}
\pgfpathlineto{\pgfpoint{\pgf@xa}{\pgf@yb}}
\pgfpathlineto{\pgfpoint{\pgf@xc}{\pgf@yb}}
\pgfpathlineto{\pgfpoint{\pgf@xb}{\pgf@yc}}
\pgfpathlineto{\pgfpoint{\pgf@xb}{\pgf@ya}}
\pgfpathclose
\pgfpathmoveto{\pgfpoint{\pgf@xc}{\pgf@yb}}
\pgfpathlineto{\pgfpoint{\pgf@xc}{\pgf@yc}}
\pgfpathlineto{\pgfpoint{\pgf@xb}{\pgf@yc}}
\pgfpathlineto{\pgfpoint{\pgf@xc}{\pgf@yc}}
}
}
\makeatother

\definecolor{colorbrewer1}{RGB}{55,126,184}
\definecolor{colorbrewer2}{RGB}{228,26,28}
\definecolor{colorbrewer3}{RGB}{77,175,74}
\definecolor{colorbrewer4}{RGB}{152,78,163}
\definecolor{colorbrewer5}{RGB}{255,127,0}
\colorlet{colorbrewer6}{yellow!90!black}
\definecolor{colorbrewer7}{RGB}{166,86,40}
\definecolor{colorbrewer8}{RGB}{247,129,191}
\definecolor{colorbrewer9}{RGB}{153,153,153}
\colorlet{colorbrewer10}{teal}
\definecolor{maincolor}{RGB}{0,110,137}

\pgfplotscreateplotcyclelist{colorbrewer}{%
	colorbrewer1, thick, mark=*\\
	colorbrewer2, thick, mark=*\\
	colorbrewer3, thick, mark=*\\
	colorbrewer4, thick, mark=*\\
	colorbrewer5, thick, mark=*\\
	colorbrewer6, thick, mark=*\\
	colorbrewer7, thick, mark=*\\
	colorbrewer8, thick, mark=*\\
	colorbrewer9, thick, mark=*\\
	colorbrewer10, thick, mark=*\\
}    
    
\usepgfplotslibrary{dateplot}

\usepackage[none]{hyphenat}

\begin{document}

\title{Computational Methods in Professional Communication
\thanks{This research was supported by the Digital Society research program funded by the Ministry of Culture and Science of the German State of North Rhine-Westphalia.}
}

\author{\IEEEauthorblockN{André Calero Valdez}
\IEEEauthorblockA{\textit{RWTH Aachen University}\\
Aachen, Germany \\
calero-valdez@comm.rwth-aachen.de}
\and
\IEEEauthorblockN{Lena Adam}
\IEEEauthorblockA{\textit{University of Münster}\\
Münster, Germany \\
l\_adam01@uni-muenster.de}
\and
\IEEEauthorblockN{Dennis Assenmacher}
\IEEEauthorblockA{\textit{University of Münster}\\
Münster, Germany \\
dennis.assenmacher@wi.uni-muenster.de}
\and
\IEEEauthorblockN{Laura Burbach}
\IEEEauthorblockA{\textit{RWTH Aachen University}\\
Aachen, Germany \\
burbach@comm.rwth-aachen.de}
\and
\IEEEauthorblockN{Malte Bonart}
\IEEEauthorblockA{\textit{TH Köln}\\
Cologne, Germany \\
malte.bonart@th-koeln.de}
\and
\IEEEauthorblockN{Lena Frischlich}
\IEEEauthorblockA{\textit{University of Münster}\\
Münster, Germany \\
lena.frischlich@uni-muenster.de}
\and
\IEEEauthorblockN{Philipp Schaer}
\IEEEauthorblockA{\textit{TH Köln}\\
Cologne, Germany \\
philipp.schaer@th-koeln.de}
}

\maketitle

\begin{abstract}
The digitization of the world has also led to a digitization of communication processes. Traditional research methods fall short in understanding communication in digital worlds as the scope has become too large in volume, variety, and velocity to be studied using traditional approaches. In this paper, we present computational methods and their use in public and mass communication research and how those could be adapted to professional communication research. The paper is a proposal for a panel in which the panelists, each an expert in their field, will present their current work using computational methods and will discuss transferability of these methods to professional communication.
\end{abstract}


\input{10_introduction.tex}
\input{20_computational_methods.tex}
\input{30_agentbasedmodeling.tex}
\input{40_socialnetworkanalysis.tex}
\input{50_naturallanguageprocessing.tex}

\input{60_searchengines.tex}

\input{70_conclusion.tex}
\section*{Acknowledgment}
We are thankful to Johannes Nakayama for proof-reading and editing the final document. This research was supported by the Digital Society research program funded by the Ministry of Culture and Science of the German State of North Rhine-Westphalia.

\bibliographystyle{IEEEtran}
\bibliography{references}

\section*{About the authors}

\textbf{André Calero Valdez} (PhD) is junior research group leader at the RWTH Aachen University, Germany. His research focuses the interaction of humans and algorithms such as recommender systems, information visualization, and machine learning in a wide variety of applications. The fields of application range from social media over health informatics to industry 4.0.  

\textbf{Lena Adam} is a PhD candidate at the Chair of Information Systems and Statistics, University of Muenster. She is member of a junior research group which focuses on Social Media analytics and online communication. By applying machine learning techniques, she strives to analyze and detect online propaganda and automated account behavior. 

\textbf{Dennis Assenmacher} is a PhD candidate at the Chair of Information Systems and Statistics, University of Muenster. His research focuses on supervised and unsupervised Social Media analytics, especially in context of systematic user exploitation to manipulate
public opinion. He is furthermore involved in the development of text-based stream clustering algorithms.

\textbf{Laura Burbach} (M.A.) is research assistant at the Human-Computer Interaction Center. Since 2018 she is part of the junior research group 'Digitale Mündigkeit'. She is currently investigating whether and to what extent different recommendation systems, voice assistants and individuals are accepted by users of social media. Besides her research focuses on target groups of Life-Logging.

\textbf{Malte Bonart} is a doctoral researcher working at Technische Hochschule Köln. His research focuses on the composition and evolution of query suggestions in person related web searches. Previously, he was a research assistant at the GESIS department of Computational Social Science where he collected, analyzed and visualized large amounts of textual, social media data. Malte Bonart studied Economics and Computer Science at the University of Cologne and the Distance University in Hagen.

\textbf{Lena Frischlich} (PhD) is a junior research group leader at the University of Münster, Germany. Her research focuses on online communication and the changing digital landscape. In particular she studies the staging and effects of online propaganda and related phenomena (e.g. disinformation) and strategies for empowering media users against manipulative attempts. She studies these questions from an interdisciplinary and multi-methodological perspective, combining quantitative, qualitative, and computational measures.

\textbf{Philipp Schaer} is Professor for Information Retrieval at TH Köln (University of Applied Sciences). He was team leader and postdoctoral researcher at the GESIS department Computational Social Science where he led a team of computer, social and information scientist. He published more than 50 peer-reviewed papers on information retrieval-related topics like query expansion, applied informetric methods in digital libraries, and evaluation of information systems.

\end{document}

%% file: 10_introduction.tex
\section{Introduction}
One of the drastic changes in recent years is the increase of the amount of available information on the Internet. We have arrived in the Zettabyte era in 2013 as the total amount of information available to the human race surpassed the Zettabyte mark in 2013~\cite{zettabyte}. In 2018 1.6 Zettabytes of information were transmitted over the Internet throughout the year~\cite{petschow2014cisco}. A Zettabyte corresponds to ``a one with twenty-one zeros'' characters of data. The sheer amount of data available becomes unfathomable to normal human proportions. When printed on regular paper, the size of this stack of paper would reach the moon approx. 260,000 times. With the availability of such large amounts of data---or so-called Big Data---novel methods need to be developed that should help to understand what useful information is stored inside this data. 

%% file: 20_computational_methods.tex
\section{Computational Methods}
Luckily, in the field of communication science traditional research approaches such as qualitative and quantitative empirical methods are no longer the only tools available to understand the world. In particular, the field of mass communication and public communication has established the use of computational methods as a means to understand how a fully connected public sphere organizes itself. The focus of research is on how digital mass media or algorithms shape communication processes and social interactions.  

For instance, such methods allow to study how different online platforms differ in social practices such as addressing hate speech, censorship, and incivility~\cite{rieger2018hate}. In the case of large social media platforms this becomes possible by applying methods from natural language processing on the large bodies of text. 

In addition to these content-focused approaches, digital methods allow us to understand the social structure of communicating people by applying methods from network research, such as social network analysis methods~\cite{schmid2018homophily}. These methods shed light on how structural elements of the social network influence the flow of communication, the spread of information (fake or truthful), and the spread of social practices in a network. 

While social network analysis focuses on the structure of the network, the individual person is often assumed to be a particle in a homogeneous anonymous mass. However, the individual differences play a large role in online communication behavior and this is where another form of digital methods play a large role. Agent-based models allow to simulate behavior of heterogeneous agents, which are virtual models of humans. Since each agent is modeled individually, this technique allows to understand behavior of large groups even when individual adaptations or structural changes occur~\cite{valdez2018human}. 

The last aspect of computational research methods aims at trying to understand how computerized systems influence communication by investigating large amounts of data created by such systems. A typical field of application is the study of search engine biases~\cite{bonart2018intertemporal}. While search and recommendation algorithms (in the narrow sense) are agnostic to human values they might still replicate biases that occur in the data they index or the usage information they rely on. This leads to questionable results that include sexism, racism or other forms of discrimination and can influence searchers in the way they perceive the digital and non-digital world.

%% file: 30_agentbasedmodeling.tex
\section{Agent-Based Modeling}
One of the inherently digital approaches to understanding human interaction and communication processes is \textit{agent-based modeling}. The general idea behind agent-based modeling is that with the use of computers it becomes possible to not only estimate individual behavior---as classical empirical research---but also simulate a large amount of individual behaviors and their interaction. By simulating multiple independent individuals on a micro level, macro level phenomena can be investigated. 

Classical empirical approaches try to identify associations, rules, laws, or formulas from observation on the one hand (exploratory research). On the other hand observations and experiments are used to confirm predictions made from theoretical considerations (confirmatory research). Theory is used to formulate predictions about the world and ideally, researchers then isolate individual variables that affect the outcome by carefully designing the experimental setup. By observing outcomes from differing preconditions, statistical evidence for mathematical formulations of the represented phenomena are gathered. If the data fits the generated hypotheses, the credibility of the theory is increased. Social empirical research is thus characterized by the duality of theory and empiricism. Theory is used to generate hypotheses. And observations and experiments are used to test hypotheses to confirm or reject theory. 

One of the key challenges of social empirical research is that by isolating effects and variables the complex interaction of effects is possibly lost. As long as effects are linear and additive in nature isolation does not harm the experimental paradigm. But in many cases effects are non-linear and are based on feedback loops. Many such effects are present in communication research. Famous examples are the spiral of silence and agenda setting theory. 

The spiral of silence, proposed by Noelle Neumann in the 1970s~\cite{noelle1980schweigespirale}, assumes that minority opinions do not find their way in the public discourse if they are antagonistic to the perceived majority opinion. People with such opinions self-censor opinion expression as they fear social isolation. 

 However, the question of when the opinion climate reaches a state in which such perceptions occur is dependent on the expression of opinions of the ``majority''---a feedback loop is created.

Agenda setting theory on the other hand proposed that the media not only determines the public sphere by reporting on current events but also determines the political agenda by selecting events. However, this selection process may be triggered by market dynamics, purposeful agenda setting, or random non-linear effects~\cite{bernard1963press}.

When social empirical research tries to validate hypotheses derived from such non-linear theories it can only ever rely on observational data. No social scientist can create an alternative public sphere for a large set of study participants for a larger period of time. Thus, effects in such investigations are either limited to observational evidence, lacking experimental isolation and controls, or are limited to micro-level models that can be verified in controlled experiments. Large scale non-linear effects are not experimentally controllable. 

The general idea behind \textit{agent-based modeling} is that by simulating micro-level (e.g., behavioral, affective, etc.) models on multiple individual agents that may interact with one another, systematic and non-linear affects become observable. The approach is to create computer equivalents of humans whose behavior is governed by mathematical formulas borrowed from scientific theories. Besides the individual behavior, the interaction is simulated mathematically as well. This process is also derived from theory. In conjunction of simulating individual behavior and interaction, agent-based modeling allows to model effects that are non-linear in nature. By randomizing starting conditions and observing which effects are independent from the starting conditions, which are sensitive to them, and which patterns occur in the simulation, macro-level models can be confirmed (or rejected) from micro-level theory~\cite{bonabeau2002agent,gilbert2008agent}. 

Experiments that are based on agent-based models hardly ever claim to match reality to the full extent. On a more fundamental level they claim to prove mechanistic, systemic effects of interaction patterns. Naturally, findings from such models depend on the quality of the underlying micro-level theories and their respective mathematical instantiation. However, if applied correctly, agent-based models allow to understand effects in their dynamical context and allow prediction of non-linear, emergent, and chaotic systems.

\subsection{Applications of Agent-based Modeling in Communication Research}

Agent-based models have been used in the context of communication science, whenever social interactions in communication play a role. With the rise of Web 2.0 and user-generated content in the recent years, the effect of users generating content on the web has become a focus of attention. As every user now has---at least potentially---turned into a publisher, traditional theories on media effects have to be rethought.

With the digitization of the media world, however, centralized mass media (e.g., television, radio, magazines) are increasingly losing importance. Social media (e.g., blogs, social networks, etc.) and user-generated content~\cite{kaplan2010users} are increasingly used. Anyone can become a publisher. But also classical media providers present themselves on the net. This diversification of mass media leads to an abundance of information. Whereas in the past the ``Sunday news'' was sufficient to inform oneself about world events, now various online contents have to be found, read and evaluated. In order to keep users on their websites for as long as possible (more advertising revenue), social media websites use so-called {recommender systems}. Recommendation systems~\cite{resnick1997recommender} have the goal of making the flood of information manageable by pre-selecting content that corresponds to the user's taste and reading preferences. An algorithm filters who gets to see which information. 

An assumed consequence of these recommendation systems is the so-called {filter bubble}~\cite{pariser2011filter} or the Echo-Chamber. 
Every user is only shown content that corresponds to their taste, opinion and political position and---in turn---is not confronted with the spectrum of social diversity.
First investigations~\cite{Dylko2017181} show that the filter bubble has an influence on the formation of political opinion. Even if users are more satisfied~\cite{Nguyen2014677}, the diversity of content decreases over time. This effect is additionally reinforced by preferable media consumption~\cite{Liao20132359}. People avoid information that runs counter to their convictions and at the same time is of great importance to them. 
In an analysis of the surfing behavior of 50,000 Americans, it was shown that segregation effects arise in the use of social media, although various topics are presented. 

Agent-based models have been used to investigate these phenomena on a structural level. Bessi et al.~\cite{Bessi20162047} found that homophily alone---the desire to associate with people that are similar to oneself---may explain the spread of misinformation in social media. When users are exposed to a variety of diverse opinions on social media, users may pick information that is in line with their beliefs and the beliefs of their friends~\cite{Bakshy20151130}.
Dandekar et al.~\cite{Dandekar20135791} found that the process of assimilation of news messages, with the use of homophily in a simulation causes polarization in its recipients. Even more disheartening, Maes \& Flache~\cite{Maes2013} found that such differentiation may even occur without prior tendency in distancing of participants.

\subsection{Applicability for Professional Communication}
Similarly to the early agent-based models by Schelling~\cite{schelling1971dynamic}, such models are often used to study the effect of an individual tendency (e.g., homophily) on system behavior (e.g., segregation). These models easily translate to communication research~\cite{Rivera2006126}. 

Agent-based models are thus very well suited to investigate the effect of communication in groups. The prerequisite is, that models for individual behavior or cognition exist on which agent-based approaches can be applied. 

Flache \& Maes~\cite{Flache200823} applied this principle in order to understand how timing affects team cohesion in demographically diverse teams and found that it is helpful to keep teams separated in homogeneous sub-groups until sub-group consensuses are reached. Only then should larger groups be formed.

Calero Valdez et al.~\cite{valdez2019predicting} used collaboration data from team output (i.e., co-authorships on documents) to determine the social infrastructure of a group. They further used these data to build an agent-based model to predict acceptance of new technology introduced to the whole group. They were able to predict acceptance mostly from the infrastructure of collaboration and few key people in the group.

Applying agent-based models to team communication processes may help in determining key opinion leaders that need to be leveraged to reach the whole group. Ignoring such dynamics may lead to ineffective communication strategies. In many such cases though, it is necessary to identify the opinion leaders from a large network of people. This is were social network analysis methods are used. 

%% file: 40_socialnetworkanalysis.tex
\section{Social Network Analysis}
Methods from social network analysis are based on the idea that we can understand the underlying social structure by looking at the structure of referents to the social structure. For example, when we look at who talks how frequently with whom, we can understand who is important in a group of people. This type of study was famously conducted by J.L. Moreno in his sociometric studies~\cite{moreno1943sociometry} in the early 1940s. The early ideas of sociometry were based in manually drawing connections between people and visually determining people that are relevant to the network. This rather qualitative approach has transformed into a quantitative science with the advent of modern computers. 

Through the use of graph theory, a subdiscipline in discrete mathematics, we have a formal definition of a network that can be translated to a computer-based transcription. Networks or \textit{graphs} are constructed using two sets ($G=(V,E)$). The \textit{nodes} are the individual entities that we look at, e.g., people in a group. Nodes are referred to as vertices in discrete mathematics and enumerated using a subscript notation ($v_i$). The set of all vertices or nodes is denoted as $V$. Connections between entities, such as friendship relations in a group, are called links or edges and are in noted as tuples of edges (e.g., $e_1 = (v_1, v_2) \in E$). The set of all edges is denoted as $E$. 

Edges may be directed---for example connecting $v_1$ to $v_2$---or undirected. In the latter case, relationships are considered to be symmetrical. Not all real relationships can be assumed to be symmetrical. Love is a relationship that is not necessarily, but ideally, symmetrical. Other relationships are necessarily symmetrical, such as co-authorship on journal papers. 
In a weighted graph, individual edges may be assigned a weight value ($w$) indicating a numerical referent for the strength of the relationship (e.g., in co-authorship how many papers were written together).

Representations of networks are typically either edge and node lists, or so-called adjacency matrices. An adjacency matrix is created by creating a $n \times n$ matrix, where $n$ is the amount of vertices (how many people). All entries of this matrix are set to zero, only when an edge exists between two vertices (e.g., $e_1 = (v_1,v_2)$) we fill the entries that match the row and column index with a one or the weight of the edge. The following adjacency matrix $A$ represents a graph of three nodes where node one and two are symmetrically connected. Or when translated to real social meaning it could be a group of 3 people, where two are mutual friends.
 \[
   A=
  \left( {\begin{array}{ccc}
   0 & 1 & 0\\
   1 & 0 & 0\\
   0 & 0 & 0 \\
  \end{array} } \right)
\]

Many real life data that is available digitally can be transformed to adjacency matrices. For example, email sender/recipient relationships, project collaboration, scientific authorships, twitter followers, or trust relationships have been modeled as graphs. The benefit of such representations is that a set of effective algorithms exist that can determine interesting properties of the graph. This allows us to determine key questions about the structure of the graph and thus also about the underlying relationships that were modeled. 

For the sake of simplicity and without loss of generality, we assume that the vertices of a graph would be people and the edges would describe whether people are on talking terms (e.g., colleagues, friends, etc.).
We can determine the \textit{radius of a graph} which indicates the longest possible path from one person to another~\cite{kang2011centralities}. This can help determine, how many steps information must take at most to reach everyone in the network.
A famous result from this approach is the small-world property of networks with high-clustering coefficients---explained further down. In small world networks every vertex is reached in only a few steps. Acquaintanceship is a relationship in real-life that yields a small-world network. It may take at most 7 steps to reach any person on the world~\cite{backstrom2012four}, if you knew everyone's acquaintances. Facebook has determined that it only takes 3.57 steps on average to reach everyone else on facebook~\cite{edunov2016three}.

We can effectively determine the \textit{shortest paths} between two vertices~\cite{dijkstra1959note}. A path is an ordered list of edges, where following items share one vertex (e.g., $ a = [(v_1, v_2),(v_2, v_3)]$). This allows to determine, who one would need to talk to, to reach a certain individual on a network. This does not only allow us to know how many individuals I must reach out to, to reach everyone, but also whom.

The \textit{centrality} of a vertex can be determined using multiple algorithms~\cite{newman2001structure}. It measures the influence of a vertex on the whole network by measuring the influence of all sub-networks and using those as a weight for influence. Translated to real life personal influence it means that if someone has influence over me, the value of this influence is larger, when I have influence over more people myself. Different measures of centrality can be obtained depending on the algorithm in use. Frequently used approaches include \textit{eigenvector centrality}, \textit{betweenness centrality}, \textit{closeness centrality}, and \textit{Katz centrality}. Each measures depicts a different property of the network.
Closeness centrality measures how close a vertex is to all other nodes in the network. Betweenness centrality measures how many of all shortest paths in the graph go through a vector. Eigenvector and Katz centrality are two different flavors of measuring influence in a network.

\begin{figure}
    \centering
    \includegraphics[width=0.8\columnwidth]{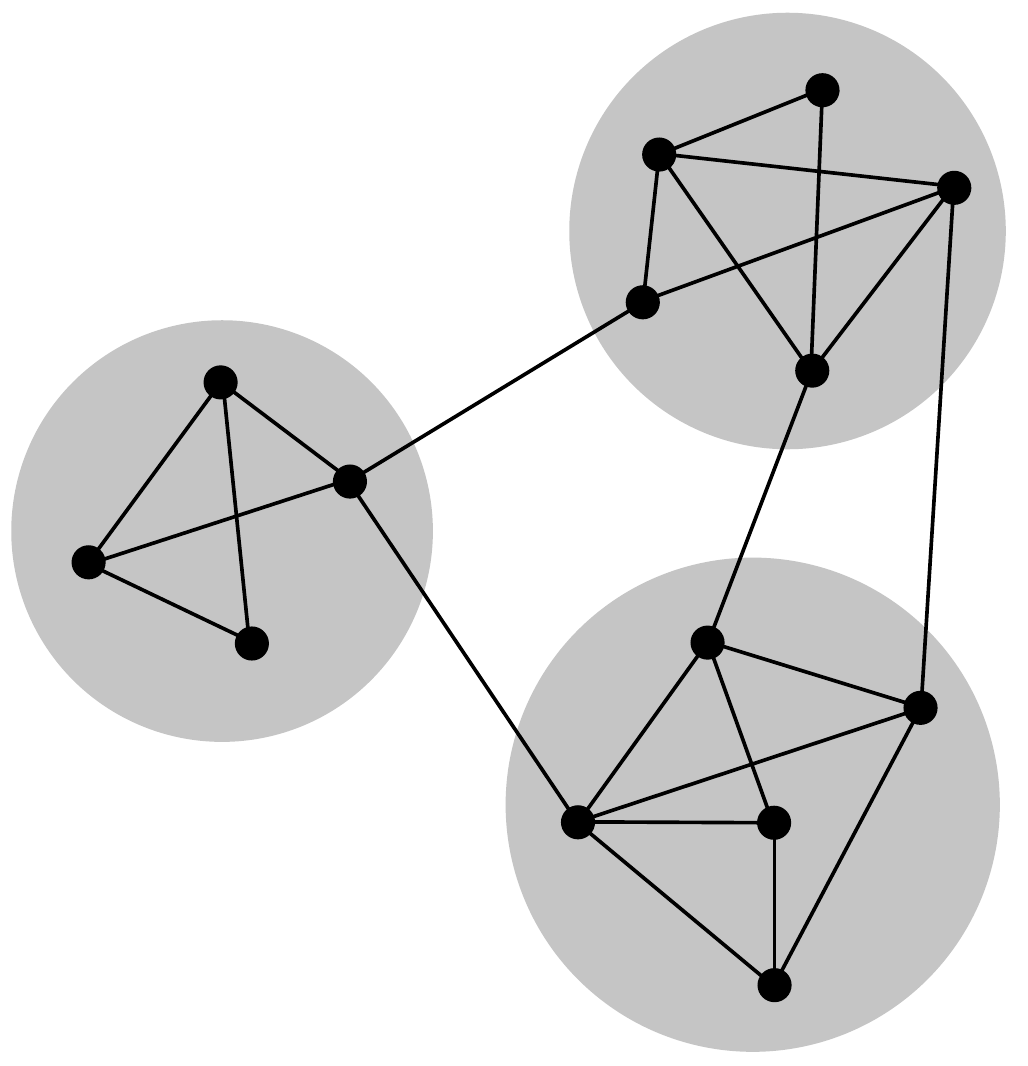}
    \caption{Example of communities in a graph. Figure used under CC-BY-SA, Copyright by J ham3 (source: wikipedia).}
    \label{fig:communities}
\end{figure}

In larger networks we can determine \textit{probabilistic clusters} or \textit{detect communities}~\cite{girvan2002community}. Several algorithms exist to determine whether a set of vertices are more closely related to each other than to the rest of the network (see Fig.~\ref{fig:communities} for an example). Key differences in those algorithms relate to whether a node must belong to a single community, may belong to multiple communities, or only has probabilistic belonging. Some algorithms (i.e., generative ones) are even able to predict missing edges during runtime.

\subsection{Applications of Social Network Analysis in Communication Research}
Methods from social network analysis have been used widely in many areas of scientific research. They have been used to identify opinion leaders in social networks such as twitter~\cite{maharani2014degree,xu2014predicting} using the individual users as vertices and the follower relationship as edges.

Such methods have been used in scientometrics---the quantitative study of scientific output---to determine importance of authors, papers, and journals~\cite{leydesdorff2011indicators}. Other approaches aimed at identifying research communities~\cite{blondel2008fast}. Even visualization approaches to understand the structure of the whole scientific community~\cite{perianes2016constructing} are regularly applied. Here, a multitude of relationships are considered for networks (e.g., co-authorship, citation, co-citation, bibliometric coupling, etc.). Using complex methods such as graph-based entropy, the diversity of local networks has been used to identify the degree of interdisciplinarity in research networks~\cite{valdez2016application,valdez2012using}.

\subsection{Applicability for Professional Communication}
The application of social network analysis can readily be applied to professional communication. A large set of easy to use tools are available as open-source software (e.g., Pajek~\cite{batagelj1998pajek}, Gephi~\cite{bastian2009gephi}, several packages in the R language~\cite{csardi2006igraph}). 

As one simple example, determining key influencers in a company, when launching new software or business processes can be used to ease transformation processes in a company by providing additional support to those who might most likely relay this support further on. This could be achieved by surveying interaction meta-data and using centrality measures on the resulting relationships. 

Running community detection mechanisms on customer relationship data could help identify common clusters of customers that can be addressed using a single communication strategy. Edges in such a case would represent connections between customers who have bought the same product or service.

Using community detection on word embeddings (a graph-based representation of word relatedness) can be used to identify clusters of similar written content~\cite{bogdanova2015detecting}. These networks can be derived using text-mining approaches on large bodies of text, for example from Q\&A sites used in customer support. This can help reduce effort in answering recurring questions. When we are looking at text-based approaches, a large spectrum of additional approaches becomes available---summarized under the term \textit{natural language processing}.

%% file: 50_naturallanguageprocessing.tex
\section{Natural Language Processing}
\definecolor{colorbrewer1}{RGB}{55,126,184}
\definecolor{colorbrewer2}{RGB}{228,26,28}
\definecolor{colorbrewer3}{RGB}{77,175,74}
\definecolor{colorbrewer4}{RGB}{152,78,163}
\definecolor{colorbrewer5}{RGB}{255,127,0}

\begin{figure*}[t]
\centering
\begin{tikzpicture}
\node [draw,dashed,rectangle,minimum width=9cm,minimum height=4.8cm,label=On-Line] {};

\node [xshift=2cm,yshift=-0.2cm,draw,rectangle,minimum width=4cm,minimum height=3cm,label=Micro-Clusters](micro) {};

\node [dashed, xshift=8cm,draw,rectangle,minimum width=5cm,minimum height=4.8cm,label=Off-Line] {};

\node [xshift=8cm,yshift=-0.2cm,draw,rectangle,minimum width=4cm,minimum height=3cm,label=Macro-Clusters](macro) {};

\draw[fill=colorbrewer1,draw=colorbrewer1]([shift={(-0.1,0.6)}]micro) circle (8pt);

\draw[fill=colorbrewer2,draw=colorbrewer2]([shift={(0.7,0.3)}]micro) circle (10pt);

\draw[dashed,fill=colorbrewer3,draw=colorbrewer3]([shift={(-0.8,-0.6)}]micro) circle (10pt);

\draw[fill=colorbrewer4,draw=colorbrewer4]([shift={(0.6,-0.8)}]micro) circle (5pt);


\draw[fill=colorbrewer1,draw=colorbrewer1]([shift={(-0.1,0.6)}]macro) circle (8pt);

\draw[fill=colorbrewer2,draw=colorbrewer2]([shift={(0.7,0.3)}]macro) circle (10pt);

\draw[dashed,fill=colorbrewer3,draw=colorbrewer3]([shift={(-0.8,-0.6)}]macro) circle (10pt);


\draw[draw=black]([shift={(0.3,0.45)}]macro) circle (23pt);

\draw[draw=black]([shift={(-0.8,-0.6)}]macro) circle (15pt);

\draw [->] (micro) -- (macro) node[midway,above] {aggregate};

\node [xshift=-2cm,yshift=-1cm,draw,rectangle,minimum width=3cm,minimum height=1cm,label=center:n-grams](ngrams) {};

\node[shape=document,draw,fill=white,line width=1pt,text width=0.6cm,minimum height=1.2cm,xshift = -3.8cm,yshift=1cm](docleft) {};
\node[shape=document,draw,fill=white,line width=1pt,text width=0.6cm,minimum height=1.2cm,xshift = -2.5cm,yshift=1cm] {};
\node[shape=document,draw,fill=white,line width=1pt,text width=0.6cm,minimum height=1.2cm,xshift = -1.2cm,yshift=1cm](doc) {};


\draw [->] (doc) -- ([xshift=2.3em,yshift=1.5em] ngrams.center) node[midway,left] {aggregate};

\draw [->] (ngrams.east) -- ([yshift=-2.3em] micro.west) node[midway,yshift=-2.7em] {calculate similarities};

\draw [->,dashed] ([yshift=2.3em]docleft.west) -- ([yshift=2.3em] doc.east) node[midway,yshift=1em] {time};
\end{tikzpicture}
\caption{TextClust process.}
\label{Figure:textclust}
\end{figure*}
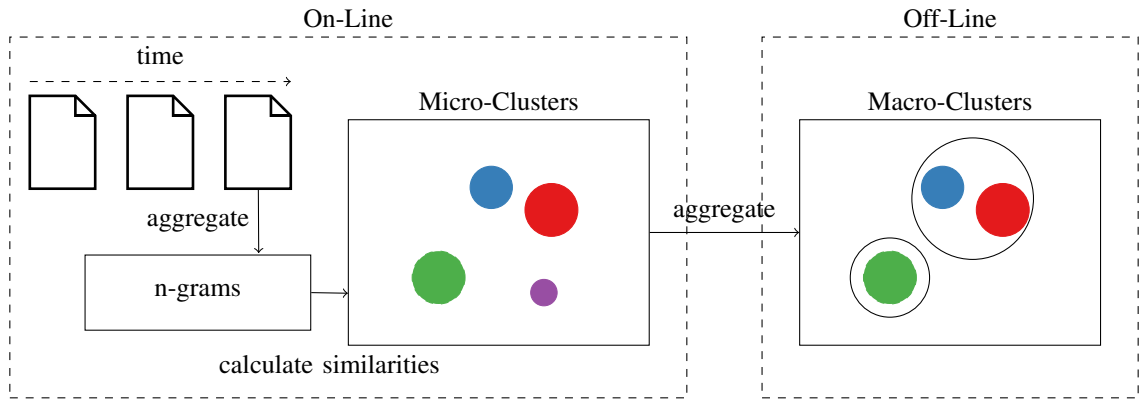

An important aspect of analyzing natural language is the so-called topic modeling approach. Topic modeling is used to automatically extract semantic meaning as well as relationships between documents from a given data corpus. Therefore, similarities between different documents are identified. In a second step, documents can be grouped together on basis of the related topics. A well-reported and most frequently used algorithm in the field of Topic Modeling is Latent Dirichlet Allocation (\texttt{LDA})~\cite{lda}. Within the \texttt{LDA} algorithm, a generative statistical model is used to describe documents as a set of latent topics. Each topic follows a unique word distribution, which in turn is generated by maximizing the conditional probability for a word to occur in a given topic.

Apart from other aspects, one major drawback of traditional topic modeling approaches (including \texttt{LDA}) is that most of the algorithms are designed for static data, i.e., fixed size data sets. In terms of communication, e.g., online discourses on social media platforms, textual data is time-dependent and potentially unbounded (data stream). It is obvious that, e.g., in discussions, topics can change over time (concept drift~\cite{BISE19}). A promising approach to detect topics and topic shifts in time-related textual data is the \texttt{textClust} algorithm~\cite{textClust}. 

\subsection{Stream Clustering as Natural Language Processing Technique}
The \texttt{textClust} algorithm is an unsupervised clustering technique that works on data streams. Clustering in general aims to identify homogeneous groups of data points. In contrast to supervised learning techniques, the number of topics, as well as the content is not known beforehand. In order to deal with temporal changes as well as with large data sets, clustering algorithms can be extended. Stream clustering methods expand upon traditional clustering concepts and solve different shortcomings of these approaches. On the one hand, traditional clustering algorithms have to iterate over the data multiple times in order to calculate similarities and arrange observations to the best fitting cluster. This is not feasible for large and potentially unbounded data sources like they might appear in continuous text data streams of online discussions. On the other hand, stream clustering algorithms provide mechanisms to decay (and ultimately forget) clusters over time. Therefore, they are able to account for the aforementioned changes within the underlying data distribution.

In general, the \texttt{textClust} algorithm follows a two-step approach, which is commonly used in the context of stream clustering. In a first step, the incoming text stream is summarized and represented as n-grams. Based on these n-grams, similarity scores are calculated, which result in a set of micro-clusters. The micro-clusters are considered as dense areas in data space. Periodically, micro-clusters are revisited, new ones are created and older entries, which are not updated recently, are decayed or ultimately removed. So far, the algorithm operates on-line, implying that micro-clusters are modified every time a new observation occurs. In a second step, micro-clusters can again be aggregated to a macro level. The aggregation to macro-clusters is part of the offline-step of the algorithm. Macro-clusters are only built on demand, which can be either after a number of new observations, or at any point in time if the user feels the need for it. Within Figure \ref{Figure:textclust}, the process of stream clustering is displayed. An in-depth description of the \texttt{textClust} algorithm can be inspected in~\cite{textClust}\footnote{The code can be accessed at \url{https://wiwi-gitlab.uni-muenster.de/stream/textClust}}.

\subsection{Applicability for Professional Communication}
 Despite it's novelty, research already exists indicating promising results when \texttt{textClust} was applied. Potential fields of application include the analysis of time-related text corpora, e.g., volatile content on websites. Assenmacher et al.~\cite{assenmacher2019openbots} examined the development of shared social bot code repositories on code-sharing platforms. They utilized \texttt{textClust} to analyze the descriptions of code-repositories and observed trends within the social bot programming community. 
 
\begin{figure*}[h]
	\centering
	\begin{tikzpicture}
	   \begin{axis}[
	   scaled y ticks=false,
		scaled x ticks=false,
			date coordinates in=x,
			width=\textwidth,
			height=5cm,
			ylabel={Cluster Weights},
			cycle list name=colorbrewer,
			enlarge y limits=.1,
			ymin=0,
			enlarge x limits=.05,%
			xticklabel style= {rotate=45,anchor=north east},
			legend entries={religion,sharia,jet},
			legend columns = 3,
			legend style={at={(0.5,1)},anchor=south,draw=none,/tikz/every even column/.append style={column sep=0.25cm}},
			major grid style={dotted},
			]
			\addplot+[mark=none, line width=1pt, line legend, colorbrewer2] table [x=time, y=weights]{figures/data/textclust_gab.txt};
			\addplot+[mark=none, line width=1pt, line legend, colorbrewer3] table [x=time, y=weightsh]{figures/data/textclust_gab.txt};
			\addplot+[mark=none, line width=1pt, line legend, colorbrewer1] table [x=time, y=weightj]{figures/data/textclust_gab.txt};

		\end{axis}
	\end{tikzpicture}
	\caption{Weights of micro-clusters over time.}
	\label{Figure:topics}
	\end{figure*}
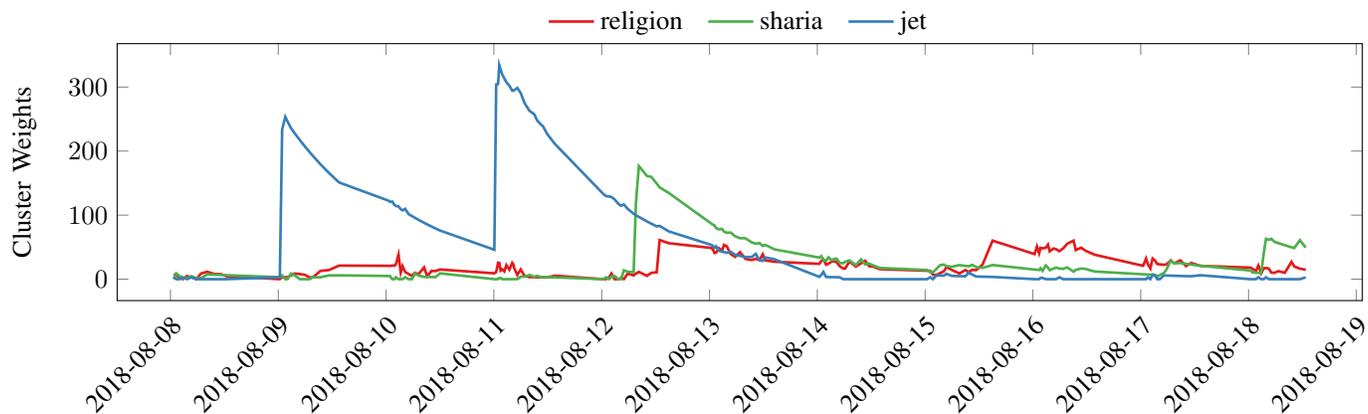

 Another use-case is the monitoring of social media platforms like Twitter, Facebook etc. There is cumulative evidence that especially alternative platforms like Gab (Twitter variant) play a crucial role for spreading extremist propaganda, disinformation, and racism, often tackling religious minorities, such as Jewish or Muslim fellow citizens~\cite{Zannettou2018a}. For this work, we gathered posts on the Gab platform, related to the search term \texttt{Islam} over a time span of 10 days.
 Within Figure \ref{Figure:topics}, the development of three micro-cluster weights (importance) are displayed. 
 While the topic religion is constantly discussed, shifting between weights of 0 and 80, two other topics with large importance peaks occurred during the recorded time span. 
 
 The topic jet (from jet plane) is not present at the beginning, increases radically at October 9 and again at October 11, and decreases to 0 afterwards. Having a look into the corresponding observations, reveals, that most of the posts are re-posts and basically contain the same content. Within these posts, the users debate about the technical functioning of airplane engines. Furthermore, we found no incident in this time which could be related to the importance peak of this topic. 
 
 The topic Sharia (Islamic law) is constantly discussed within the time-span of the 10 days. Nevertheless, the weight increases rapidly after October 12. Having a look at the data, again a similar situation as for the jet plane topic can be observed. The topic is 'pushed' by a small number of posts with the same content and their re-posts. It is conceivable that the topics Sharia and jet are part of an (automated) campaign.

Next to the application on social media platforms, \texttt{textClust} can be used within comment or product rating sections on web presences of enterprises. By tracking topics within customer-generated content, weaknesses concerning products or services could be detected and fixed, preventing negative publicity and increasing customer satisfaction. Furthermore, the monitoring of events, like elections, within online communication (e.g., via Twitter), is a possible use case for stream clustering. 

Our results show the potential of stream clustering, specifically \texttt{textClust} within the field of communication. Due to the large amount of volatile text content that is shared over the Internet nowadays, it becomes increasingly challenging to extract semantic meaning from the data. By providing an aggregated time-sensitive view, we overcome these challenges and make the overwhelming amount of content more accessible for end users.

%% file: 60_searchengines.tex
\section{Search engine studies}
Another typical digital approach to data is the use of indexed or non-indexed search engines. In this section, we show how the computational method of data access itself can be studied as well.

Web search engines are the entry point to the myriads of web documents available on the World Wide Web. They enable users to gain access to these documents by using everything from everyday language to highly specialized domain-specific languages. User-friendliness and high-quality results due to content-based document ranking are the key features of these modern search engines that enables access to a very heterogeneous set of documents. 
All these features make web search an everyday task and web search engines a highly trusted information source. In the 2017 Global Edelman Trust Barometer~\cite{antoine_harary_2017_2017}, web search engines were ranked as the most trusted information with an agreement rate of 64\%. 

For professional communication this introduces the question what and how information on companies, products or company representatives is presented by search engines. As much as recent works focus on social networks~\cite{roundtree_engineers_2018}, search engines and their role as intermediates are somehow neglected in the recent past. 

\subsection{Foundations and Current Issues in Search Engine Studies}

The three underlying building blocks of a search engine are the crawling, indexing and query process. First, the web is systematically scanned for new or updated websites, the second process processes and stores the websites and relevant metadata in an efficient manner. This creates the infrastructure which makes searching the web possible. The third process utilizes the infrastructure to retrieve and rank documents that are relevant to a user submitted query~\cite{croft_search_2009}. Each submitted query and many user actions such as browsing, clicking or reformulations are logged. The log is used to evaluate the retrieval and ranking processes as well as to provide query suggestions. These suggestions are presented to the users during the formulation of the query. They appear as a ranked list below the search interface and typically contain $5$ to $10$ items. Search suggestions are based on the past popularity of search queries, the location and language settings of the user~\cite{google2018}. 

Search engine result pages and search suggestions are the two main functionalities the users interact with. Current research in information retrieval is concerned with biases, unfairness and missing transparency in the retrieval and ranking mechanisms~\cite{castillo_fairness_2019, baeza-yates_bias_2018}. Bias describes the systematic and unfair exclusion, inclusion or prominence of certain items during the automated ranking process~\cite{mowshowitz_bias_2002, introna_shaping_2000}. As an example consider the case of inherent gender biases in Google's image search~\cite{kay_unequal_2015, otterbacher_investigating_2018}: In these studies the authors assessed image search results of various occupations and their perception by users. Other studies, e.g., for the case of vaccination related searches~\cite{allam_impact_2014} or searches for political candidates~\cite{epstein_suppressing_2017}, discuss the influence of manipulated rankings to the user's opinion formation and decision making. While some research studies algorithmic ways to identify and adjust for underlying misbehavior of the technical systems, e.g., through the identification of inappropriate query suggestions~\cite{gupta_learning_2017}, others tackle these problems by increasing the transparency of ranking processes and the credibility of online documents~\cite{fuhr_information_2018, yang_nutritional_2018}. 

\subsection{Long-term Monitoring of Search Engine Query Suggestions}

The monitoring and auditing of search engines for studies of bias, personalization and fairness, requires the collection of sufficient data on search engine result pages and query suggestions. Some researchers use browser extensions which are installed on personal computers of study participants~\cite{robertson_auditing_2018}. Others rely on web-scraping methods as described above~\cite{hannak_measuring_2013}. To study biases and misinformation in query suggestions in searches for political candidates, we have gathered the suggestions for the names of over $2000$ German politicians over a period of $2$ years. The search list includes the names of all members of the German ``Bundestag'' and important party and governmental leaders. It captures the period before and after the election in September 2017. For the Google, Bing and DuckDuckGo search engines, the scraping takes place twice a day so that changes in the list of suggestions can be accurately dated. Proxy servers which route the Internet traffic over computers in other countries, are utilized to control for the location and language of the search request. The suggestions are stored in a remote database such that subsets of the data can be queried efficiently. 

Especially for popular personalities, the list of suggestions is quite dynamic and often contains references towards current events. It stands out that these reference are not always related to serious headlines but can contain descriptors of rumors, misinformation or sensational news as they often appear in the yellow press. Following these suggestions sometimes lead to result pages with an overall bad quality. As the retrieval systems behind the query predictions are mainly driven by popularity measures, no credibility assessment of submitted queries takes place. A better understanding of why some news or rumors trigger enough searches such that they appear as trending suggestions can help to improve the system and quantify the credibility of single event-related suggestions. 

\subsection{Applicability for Professional Communication}

\begin{table}[htbp]
\caption{Top ten query predictions for the names of male and female business persons. Numbers measure the fraction of visible days (in \%) averaged over all query terms in the respective list.} 
\begin{center}
\begin{tabular}{llll}
  \hline
\multicolumn{2}{p{0.4\linewidth}}{$50$ ``most powerful women'' in US business \cite{fortune2018}, US region settings} & \multicolumn{2}{p{0.4\linewidth}}{$30$ German DAX, male CEO's, German region settings}  \\ 
  \hline
salary & 70 & gehalt (salary) & 56 \\ 
  linkedin & 64 & kinder (children) & 32 \\ 
  net worth & 57 & vermögen (wealth) & 28 \\ 
  husband & 55 & linkedin & 27 \\ 
  age & 34 & ehefrau (wife)& 24 \\ 
  family & 32 & frau (woman/wife) & 22 \\ 
  twitter & 32 & familie (family) & 19 \\ 
  email & 25 & interview & 18 \\ 
  house & 25 & lebenslauf (vita) & 14 \\ 
  quotes & 19 & net worth & 12 \\ 
   \hline
\end{tabular}
\label{tab1}
\end{center}
\label{tab:search}
\end{table}

In today's digitized world, finding relevant information online becomes an everyday task and for most people search engines are the entry point to the web. Therefore, they take the role of intermediate players and function as gatekeepers to the web's content. They organize the access to information and influence the way it is publicly communicated. Therefore, it can be essential for professional communicators to understand the mechanisms behind web search. In professional communication, some focus was placed to practical lessons in search engine optimization, with the goal of achieving higher ranks for the company's website~\cite{killoran_how_2013}. Other works within professional communication pick up parts of web search engine technologies, like crawling and scraping.  
Scrapers are a key component within modern search engines that extract information from web documents and allow to annotate different entities and are known to work well for professional communication~\cite{lauer_hand_2018}. Communication also manifests itself in other forms as the following example shows:

Over a period of two month from January, 23 until March, 23 2019, we have monitored the US-based Google suggestions for Jeff Bezos, the current CEO of ``Amazon''. The general suggestions \emph{net worth, house, salary, family, age, twitter} and \emph{wife} were also prevalent in most suggestions for other business persons, as illustrated in table \ref{tab:search}. But in particular for this case, the suggestion algorithm was picking up the current news around the CEO's affair: \emph{new girl, lauren sanchez, divorce court, text photos, pictures, photos, national enquirer} and \emph{girlfriend 2019}. Therefore when searching for ``Jeff Bezos'' during this time, a very distinct picture of the CEO was communicated via the drop down list of search predictions.

%% file: 70_conclusion.tex
\section{Conclusion}
The four different types of computational methods shown in this paper have proven successful in the field of mass and public communication and are also promising approaches in professional communication. This is particularly true when such methods are used in combination. For example, understanding how adoption processes of communication processes in organizations work, can be modeled using social network analysis and agent-based modeling to acquire improved predictions about organizational change~\cite{valdez2019predicting}. Using text-mining and methods to analyze algorithmic biases can be used to understand how competitors position themselves or to detect emerging competitors on the web. 

Still, all computational methods are faced with one key-challenge---data availability. As we have seen in the previous sections, data can often be acquired directly or from proxy-items. The problem here is that missing data is not represented in the results. Therefore, relying solely on computational methods without intensive reflection will introduce ethical and epistemological questions to the debate. Can we base decisions on these findings that have real societal impact? Do computational methods measure the bias of reality or the bias of data collection? Evaluating the use of computational methods in conjunction with solid quantitative and qualitative empirical social research can provide the means to determine the ranges of validity of computational methods and potentially yield completely novel research paradigms in professional communication.